\documentclass[twocolumn,floatfix, nofootinbib,prd,preprintnumbers,showpacs,showkeys,superscriptaddress,longbibliography]{revtex4-1}

\usepackage[mathscr]{euscript}
\usepackage{amsmath}
\usepackage{graphicx}
\usepackage{dcolumn}
\usepackage{bm}
\usepackage{epsfig}
\usepackage{amssymb,latexsym,mathrsfs}
\usepackage{graphicx}
\usepackage{color}
\usepackage{hyperref}
\usepackage{float}
\usepackage[thinlines]{easytable}
\usepackage{multirow}
\usepackage{diagbox}
\usepackage{varwidth}

\newcolumntype{M}[1]{>{\centering\arraybackslash}m{#1}}
\newcolumntype{N}{@{}m{0pt}@{}}

\usepackage{amsmath}
\newcommand*\diff{\mathop{}\!\mathrm{d}}

\hypersetup{
    colorlinks=true,
    linkcolor=red,
    citecolor=blue,
}

\usepackage{amsmath}
\usepackage{amssymb}
\usepackage[caption=false]{subfig}
\usepackage{hyperref}
\usepackage{url}
\usepackage{xcolor}
\usepackage{color}
\definecolor{amaranth}{rgb}{0.9, 0.17, 0.31}
\definecolor{purple(munsell)}{rgb}{0.62, 0.0, 0.77}
\definecolor{americanrose}{rgb}{1.0, 0.01, 0.24}
\definecolor{palatinateblue}{rgb}{0.15, 0.23, 0.89}
\definecolor{royalblue(web)}{rgb}{0.25, 0.41, 0.88}
\definecolor{hanpurple}{rgb}{0.32, 0.09, 0.98}
\definecolor{beaublue}{rgb}{0.74, 0.83, 0.9}
\definecolor{carminered}{rgb}{1.0, 0.0, 0.22}
\definecolor{brightpink}{rgb}{1.0, 0.0, 0.5}
\definecolor{vividviolet}{rgb}{0.62, 0.0, 1.0}

\definecolor{electron}{rgb}{1.0, 0.67, 0.22}
\hypersetup{ linktoc=all,
    colorlinks, linkcolor={palatinateblue},
    citecolor={brightpink}, urlcolor={amaranth}}

\newcommand{\be}{\begin{equation}}
\newcommand{\ee}{\end{equation}}
\newcommand{\bs}{\begin{split}} 
\newcommand{\bea}{\begin{eqnarray}}
\newcommand{\eea}{\end{eqnarray}}

\newcommand{\al}{\alpha} 
 
\newcommand{\kp}{\kappa} 
\newcommand{\gam}{\gamma}

\newcommand{\bes}{\begin{subequations}}
\newcommand{\ees}{\end{subequations}}

\newcommand{\bo}{\raise-1mm\hbox{\Large$\Box$}}

\newcommand{\bd}{\boldsymbol}

\usepackage{physics}
\usepackage{csquotes}
\usepackage{mathtools}

\begin{document}

\title{IR-finite thermal acceleration radiation} 

\author{Evgenii Ievlev}
\email{evgenii.ievlev@nu.edu.kz}
\altaffiliation[On leave of absence from: ]{National Research Center “Kurchatov Institute”, Petersburg Nuclear Physics
Institute, St.\;Petersburg 188300, Russia}
\affiliation{Physics Department, Nazarbayev University,\\
Astana 010000, Qazaqstan}
\affiliation{Energetic Cosmos Laboratory, Nazarbayev University,\\
Astana 010000, Qazaqstan}
\affiliation{Theoretical and Nuclear Physics Department, al-Farabi Qazaq National University,\\ 
Almaty 050040, Qazaqstan}
\author{Michael R.R. Good}
\email{michael.good@nu.edu.kz}
\affiliation{Physics Department, Nazarbayev University,\\
Astana 010000, Qazaqstan}
\affiliation{Energetic Cosmos Laboratory, Nazarbayev University,\\
Astana 010000, Qazaqstan}
\affiliation{Leung Center for Cosmology and Particle Astrophysics,
National Taiwan University,\\ Taipei 10617, Taiwan}
\author{Eric V. Linder}
\email{evlinder@lbl.gov}
\affiliation{Energetic Cosmos Laboratory, Nazarbayev University,\\
Astana 010000, Qazaqstan}
\affiliation{Berkeley Center for Cosmological Physics \& Berkeley Lab, University of California, 
\\Berkeley 94720, CA, USA 
}

\begin{abstract} 
A charge accelerating in a straight line following the Schwarzschild-Planck moving mirror motion emits thermal radiation for a finite period. Such a mirror motion demonstrates quantum purity and serves as a direct analogy of a black hole with unitary evolution and complete evaporation. Extending the analog to classical electron motion, we derive the emission spectrum, power radiated, and finite total energy and particle count, with particular attention to the thermal radiation limit. This potentially opens the possibility of a laboratory analog of black hole evaporation.  
\end{abstract} 

\pacs{41.60.-m (Radiation by moving charges), 05.70.-a (Thermodynamics), 04.70.Dy (Quantum aspects of black holes)}
\date{\today} 

\maketitle


\section{Introduction}

The precise metric solution to Einstein's field equations in general relativity, unveiled by Schwarzschild \cite{Schwarzschild:1916uq}, implied the existence of gravitationally collapsed stars that emit no light. Hawking proposed a semi-classical approximation \cite{Hawking:1974rv}, using a quantum field subject to the classical background of the Schwarzschild metric, to show that black holes radiate with a thermal spectrum and temperature proportional to their surface gravity.

However, this discovery presented an obstacle, as the projected number of particles and radiated energy was infinite \cite{Hawking:1974sw}, and Hawking's computation scaled distances smaller than the Planck length \cite{Jacobson:1991gr}, where the semi-classical approximation fails due to backreaction \cite{Fabbri}, known as the trans-Planckian problem \cite{Brandenberger:2010bpq}.  It appears unlikely that any direct observational confirmation of black hole evaporation will emerge \cite{Wald:1999vt}, given the weakness of the anticipated amplitude relative to the interference from cosmic microwave background \cite{Parker:2009uva}. Consequently, to date  there is no astrophysical evidence for Hawking radiation.

A semi-classical approximation, using a quantum field subject to the classical trajectory of an accelerated mirror \cite{Davies:1976hi} was also proposed by Davies-Fulling. This perfectly reflecting boundary \cite{DeWitt:1975ys} propagates through flat spacetime, a dynamical Casimir effect, radiating with a thermal spectrum and temperature proportional to acceleration \cite{Davies:1977yv} in analogy to black hole radiation. A Schwarzschild mirror trajectory \cite{Good:2016oey} exactly recreates the radiation emitted by a null shell collapse to a Schwarzschild black hole \cite{wilczek1993quantum}, including the aforementioned difficulties with infinite energy and particles \cite{Good:2018zmx,Cong:2018vqx,Good_2017Reflections1}.  A modified Schwarzschild trajectory for an accelerated mirror \cite{Good:2019tnf} resolves the infinite radiation challenges of the Schwarzschild trajectory by limiting the acceleration via the introduction of a small length scale. The corresponding spacetime metric was obtained as a regularized form of the Schwarzschild spacetime called the Schwarzschild-Planck metric \cite{Good:2020fsw}; see also its closely related optical analog \cite{Moreno-Ruiz:2021qrf}. The accelerated equation of motion itself called the Schwarzschild-Planck trajectory is a globally defined, timelike, asymptotically zero velocity straight-line path in Minkowski spacetime.   

The flat-spacetime simplicity and lower dimensions of the moving mirror model facilitate tractability in finding analytic Bogolyubov coefficients to investigate quantum particle creation \cite{Birrell:1982ix}, as well as other quantities of interest like entanglement entropy \cite{Reyes:2021npy,Akal:2020twv,Holzhey:1994we,Lee:2019adw,Bianchi:2014qua}, particle count \cite{walker1985particle}, and radiated energy \cite{Walker:1984ya}.  In fact, Ford-Vilenkin \cite{Ford:1982ct}, Nikoshov-Ritus \cite{Nikishov:1995qs}, Zhakenuly et al. \cite{Zhakenuly:2021pfm}, Ritus \cite{Ritus:1999eu,Ritus:2002rq,Ritus:2003wu}, and others have exploited this simplicity to find powerful analogies between the classical radiation of an electron and the quantum radiation of a mirror.  Indeed, a functional identity \cite{Ritus:2022bph}, up to a pre-factor of $4\pi$ times the fine structure constant, exists between the electron and mirror.  Moreover, this duality \cite{Good:2022eub} has made it possible to classically derive the Planck distribution connection between acceleration and temperature for an electron's Larmor radiation \cite{Ievlev:2023inj}.  

This work aims to investigate the classical Larmor radiation emitted by an electron traveling along the Schwarzschild-Planck trajectory to gain a better understanding of the thermal emission process usually associated with uniform proper acceleration and quantum radiation of the Davies-Fulling-Unruh effect \cite{Fulling:1972md,Davies:1974th,unruh76}.  The advantages and novelty of the classical approach are the conceptual simplicity and analytic tractability of the thermal physics of radiation and the associated analog to complete black hole evaporation. 

The paper is organized as follows. In Sec.~\ref{sec:trajectory} we introduce the Schwarzschild-Planck trajectory, focusing on its dynamics. In Sec.~\ref{sec:energy} and Sec.~\ref{sec:distributions} we compute the finite energy emitted by the electron, and the spectral-power distributions. Sec.~\ref{sec:thermality} is devoted to thermality, asymptotic limits and finite particle production.  Sec.~\ref{sec:electron-mirror} reviews some background with respect to the generalized correspondence between mirrors and electrons; refining the recipe for the duality. Finally, Sec.~\ref{sec:remarks} and Sec.~\ref{sec:conclusions} end with remarks and conclusions.

\section{Schwarzschild-Planck trajectory} 
\label{sec:trajectory}
The Schwarzschild-Planck trajectory of the quantum pure mirror \cite{Good:2019tnf} is\footnote{We will mostly employ natural units, setting $\hbar = c = k_B = 1$; however then the electron's charge is a dimensionless number $e^2=4 \pi \alpha_{\textrm{fs}} \approx 0.092$.  In this section and the next, we set $e=1$ rather than $\hbar =1$. The SI quantities of vacuum magnetic permeability and the vacuum permittivity will always be set $\mu_0 = \varepsilon_0 = 1$. The metric signature is $(+,-,-,-)$, the same as Jackson \cite{Jackson:490457}. 
The polar angle $\theta$ runs from $0$ ($z>0$) to $\pi$ ($z<0$).
} 
\begin{equation}
    t(z)=-z-\frac{1}{g}\sinh 2\kp z\ , 
\label{SP_traj}
\end{equation}
where $\kp$ and $g$ are two inverse length scales or 
accelerations. This was shown to correspond to the 
Schwarzschild-Planck black hole solution that evaporates 
without a remnant and preserves unitarity \cite{Good:2020fsw}. 
The idea is 
that one length scale $\kp^{-1}$ would correspond to the 
Schwarzschild radius or black hole mass, and the other 
length scale $g^{-1}$ would be related in some way to 
the Planck scale. For our purposes here we regard 
$\kp$ and $g$ as two parameters in an accelerated electron 
motion. 

The velocity 
\be 
V(z)\equiv \dv{z}{t} =\frac{-g}{g+2\kp\cosh 2\kp z}\ , 
\label{velocity}\ee 
is zero in the limits $z\to\pm\infty$ and hence the 
motion is asymptotically static; see e.g.\  \cite{Walker_1982}, and also the two trajectories in \cite{Good:2023ncu}. This is critical for 
finite energy and particle production, and unitarity, 
and the relation between classical Larmor power and 
quantum Bogolyubov coefficients, as discussed in 
\cite{Ievlev:2023inj,Good:2023ncu}. The maximum velocity 
$V_{\rm max}=-(1+2\kp/g)^{-1}$ occurs at $z=0$ and 
approaches the speed of light for $\kp/g\ll1$.

\subsection{Accelerations} 

The proper acceleration $\al\equiv\gam^3\ddot z$, a Lorentz invariant scalar \cite{Rindler:108404}, is 
\be 
\al=\frac{-\sqrt{g\kp}\,\sinh 2\kp z}{2\,[\cosh 2\kp z+(\kp/g)\cosh^2 2\kp z]^{3/2}}\ , 
\ee 
where $\gam$ is the Lorentz boost factor. The acceleration 
vanishes asymptotically as $e^{-4\kp|z|}\sim t^{-2}$. 

Another useful quantity is the 
``local'' or peel acceleration. 
Consider the acceleration \cite{carlitz1987reflections},
\be \bar{\kappa} = \frac{v''(u)}{v'(u)} = 2 \alpha e^\eta = 2\eta'(u)\,, \label{peel}\ee
where $\eta = \tanh^{-1}V$ is the rapidity, $v=t+z$ is advanced time, $u=t-z$ is retarded time, and $\alpha$ is the proper acceleration. Eq.~(\ref{peel}) is called the `peel', whose name has precedent in studies on the asymptotic behavior of zero-rest mass fields, see e.g.\  Penrose \cite{penrose1965zero}.  

The peel acceleration function itself was perhaps first introduced as the local acceleration in \cite{carlitz1987reflections}, and known from gravitational contexts as the peeling function, e.g.\ \cite{Bianchi:2014qua,Barcelo:2010pj}.  In the electron radiation context  \cite{Ievlev:2023inj,Good:2022xin,Good:2022eub}, the peel has also been associated with thermality.  Interestingly, and surprisingly unlike the uniform proper acceleration of Unruh temperature \cite{unruh76}, the peel $\bar{\kappa}$ is the dynamic object that exhibits uniformity instead of the Lorentz invariant $\alpha$. 
 The accelerating electron (or flying mirror) emits photons distributed in a Planck distribution with radiation temperature $T=\bar{\kappa}/2\pi$, not $T=\alpha/2\pi$.

\begin{figure}[h]
\includegraphics[width=\columnwidth]{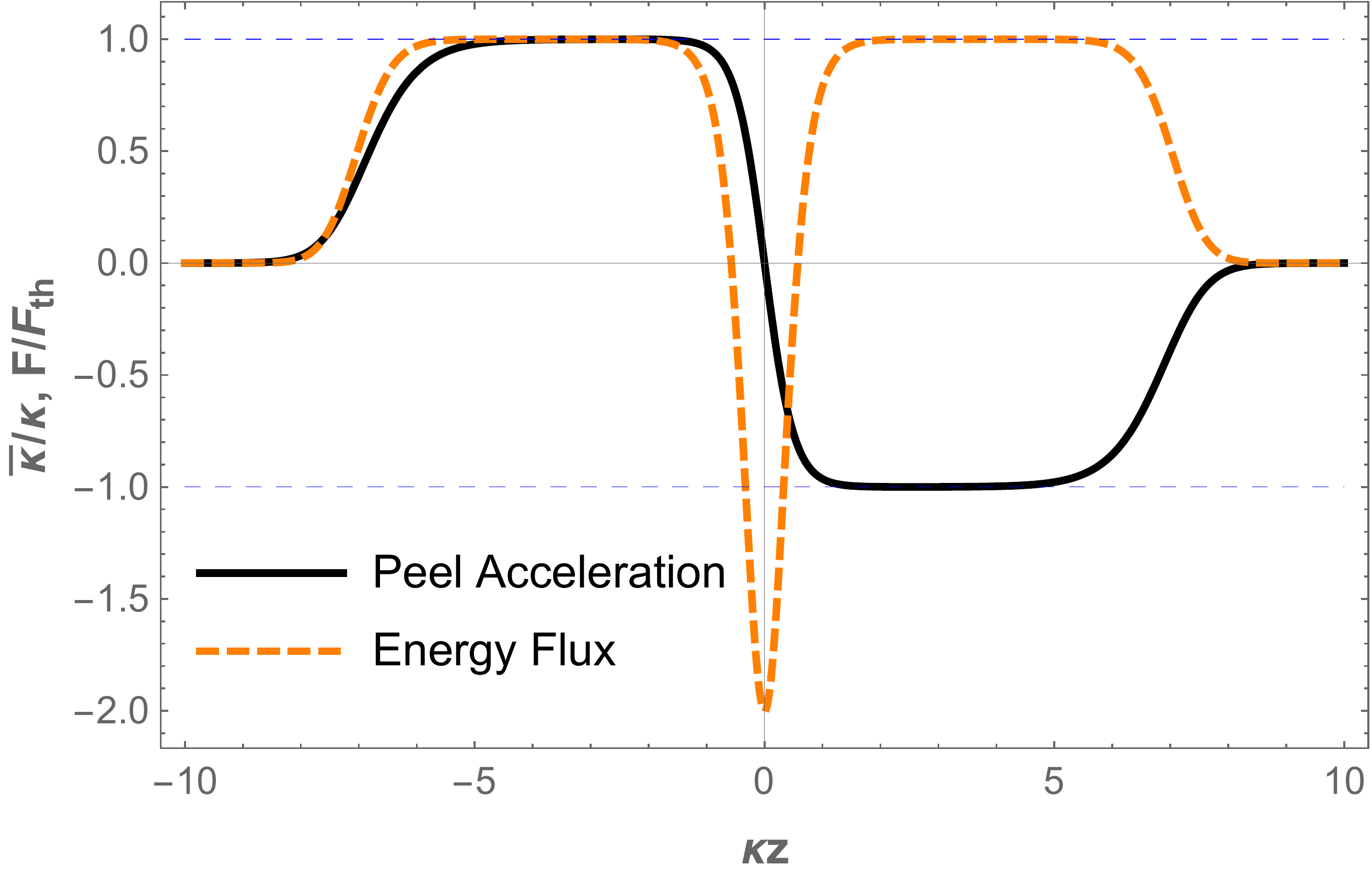} 
 \caption{A plot of the peel acceleration, $\bar{\kappa} = 2 \alpha e^\eta = v''(u)/v'(u) = 2\eta'(u)$. Here $g/\kappa = 10^6$.  A key takeaway is the leveling plateau indicative of steady-state equilibrium, which is in accordance with a Planck distribution of photons with temperature proportional to constant peel, $T= \kappa/2\pi$. For comparison, we show the energy flux, Eq.~(\ref{energyflux}),  radiated from a mirror with the same trajectory, also exhibiting a thermal plateau.
 }
\label{Fig_peel}
\end{figure}

Explicitly written using the Schwarzschild-Planck trajectory, Eq.~(\ref{SP_traj}), this object is
\be \bar{\kappa} = - \frac{\kappa \tanh 2 \kappa z}{(r \cosh 2 \kappa z+1)^2}\ ,\ee
where $ r = \kappa/g$. See Figure~\ref{Fig_peel} for a plot of this acceleration. A key feature is the horizontal leveling associated with steady-state equilibrium.  
Here we will see the peel exhibits uniformity in connection with an associated thermal Planck distribution.  The leveling out demonstrates a very slow change in the peel, commensurate with radiation emission. During this time, the system is in thermodynamic equilibrium, typical of a quasi-static process \cite{schroeder2000introduction}.

\subsection{Energy Flux and Power} 

The energy flux $F$ of emitted 
radiation for the Schwarzschild-Planck 
case of a moving mirror can have a plateau at the thermal 
value $F_{\rm th}=\kp^2/(48\pi)$, which lengthens as 
$\kp/g\to0$. Specifically, for $\kp/g\ll1$, 
\be 
F(z)\approx F_{\rm th}\,\left[1-\frac{3}{\cosh^2 2\kp z}+{\mathcal O}\left(\frac{\kp}{g}\right)\right]\ .\label{energyflux} 
\ee 
The thermal flux plateau lasts from $|\kp z|\approx[1,(1/2)\ln(g/\kp)]$. 
This corresponds to the period of 
uniform peel acceleration and 
equilibrium behavior. Note that $48\pi F(u) = \bar{\kappa}^2 - 2\bar{\kappa}'$. See \cite{Good:2019tnf} for a discussion 
of the necessary negative energy flux around $z=0$. 

\begin{figure}[h]
\centering
\includegraphics[width=\columnwidth]{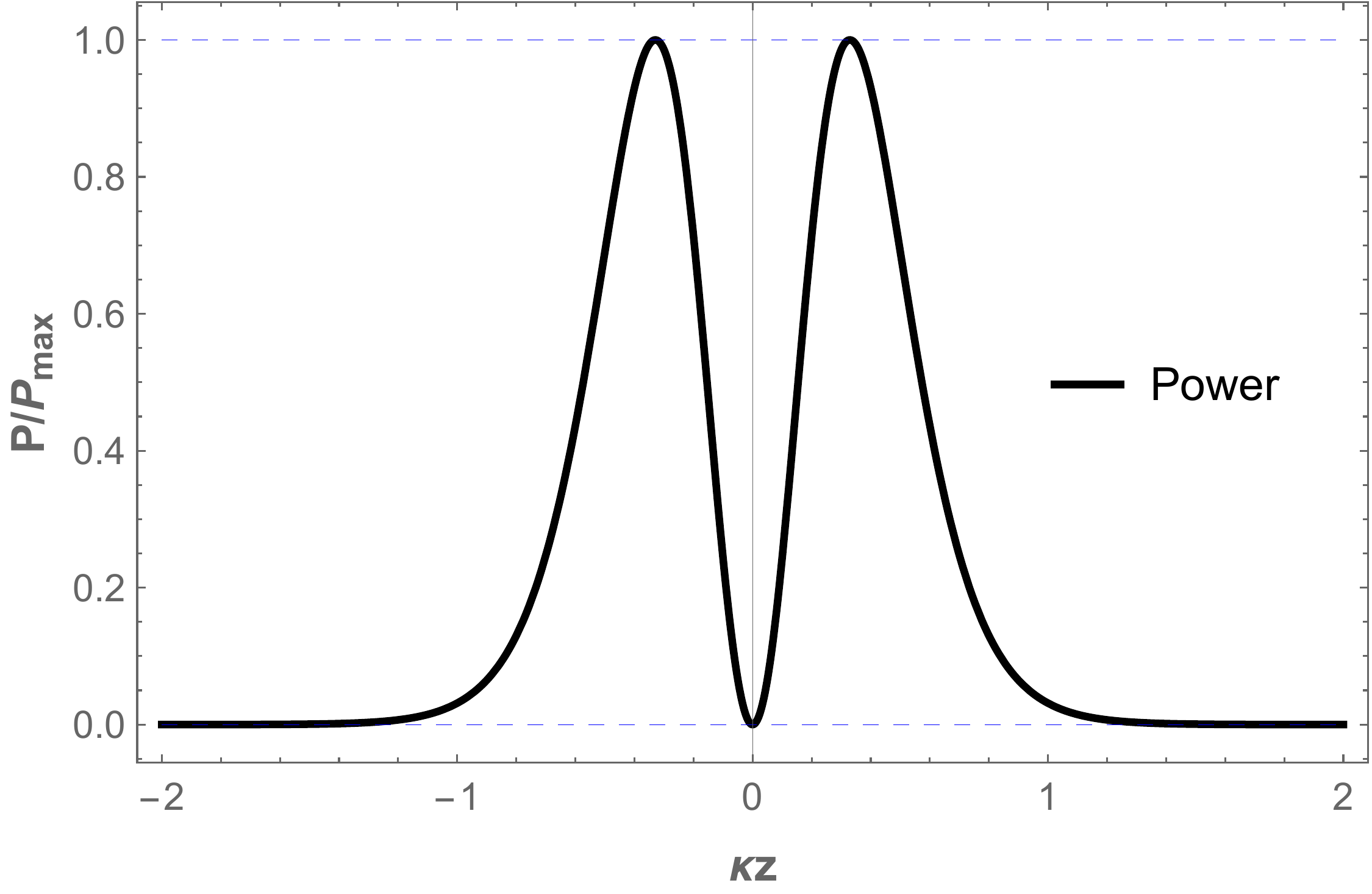}
 \caption{A plot of the Larmor power emitted by the electron, Eq.~(\ref{eq:power}), $P = \alpha^2/6\pi$, normalized by its maximum value. A key takeaway is that the power is zero at the asymptotes, and the area under the curve, the total energy emitted, is finite.  
 Here $g/\kappa = 10^6$. 
 } 
\label{Fig_power}
\end{figure}

For simplicity, allow us to set $e=1$ in this section and the next. For a radiating charge, the Larmor power
\be 
P(z)=\frac{\al^2(z)}{6\pi}\ , \label{eq:power} 
\ee 
is of interest, shown in Figure~\ref{Fig_power}. 
Note that for tractability we express the power as a function of the spatial variable $z$ rather than the time variable $t$.
Since the power dies 
off at large $z$ as $e^{-8\kp z}$, or equivalently at 
large times as $t^{-4}$, we will obtain finite total energy.

\section{Finite Energy} 
\label{sec:energy}
To compute the total energy of the quantum pure mirror (which is the same as the total energy radiated by the electron, see \cite{Ievlev:2023inj}), we can integrate the Larmor power over all time to give 
\bea   
E&=&\frac{1}{6\pi}\int_{-\infty}^\infty dz\,\frac{dt}{dz}\,\al^2\\ 
&=&\frac{g^3}{24\pi}\int_0^w dw\,\frac{\sinh^2w}{\cosh^3w}\,\frac{g+2\kp\cosh w}{(g+\kp\cosh w)^3}\ , \label{eq:energy} 
\eea 
where $w=2\kp z$, $r=\kappa/g$, and we have used the evenness
of the integrand to change the interval from $[-\infty,+\infty]$ to
$2\times[0,\infty]$. 

\begin{figure}[htbp]
\centering
\includegraphics[width=\columnwidth]{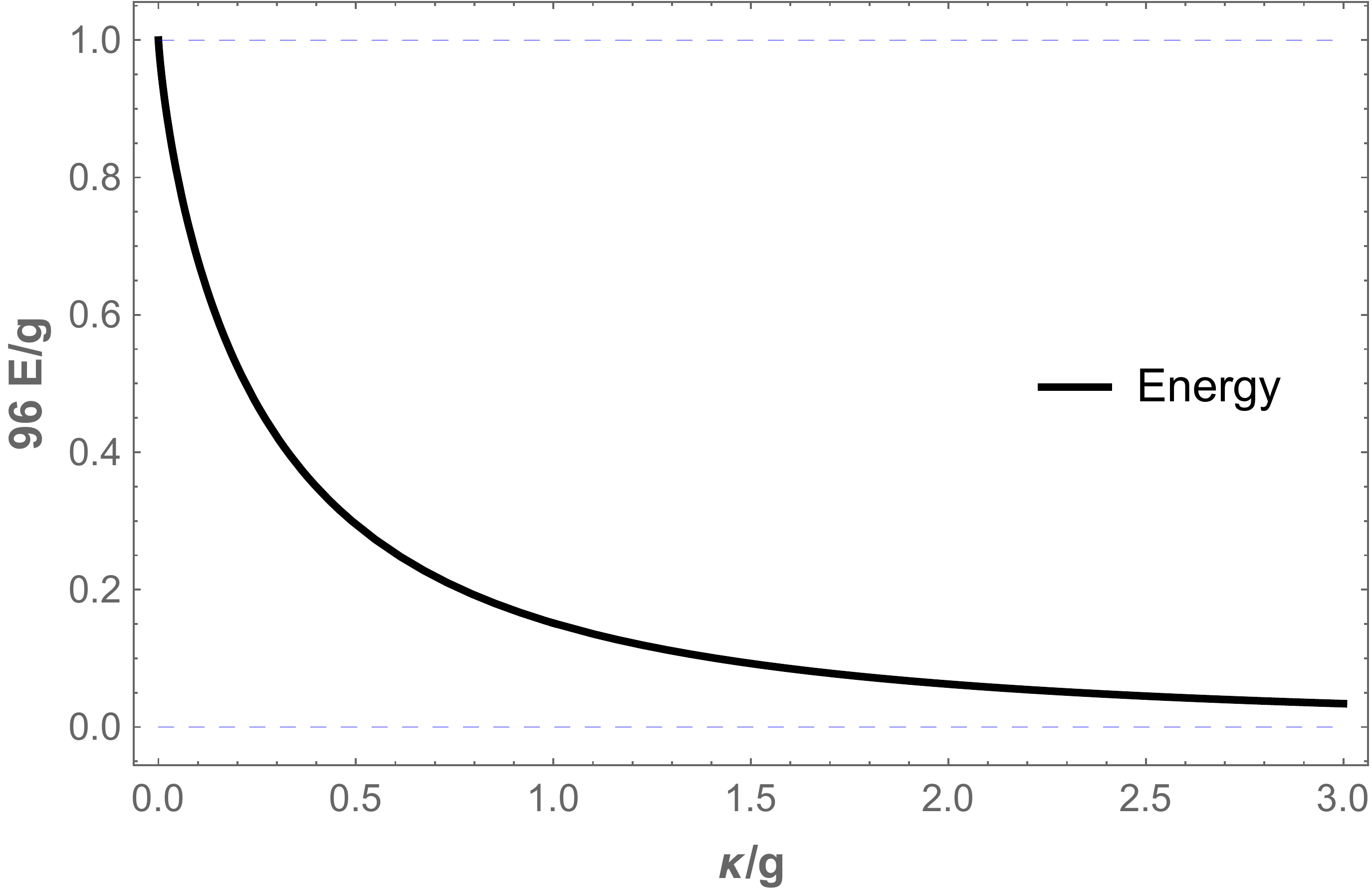}
 \caption{The total energy emitted by the electron, Eq.~(\ref{eq:energy}), is shown normalized to its value $g/96$ 
 as $\kp/g\to0$.  
Notice the thermal regime corresponding to $g\gg \kappa$ has greater total energy emission while less energy is emitted in the 
non-thermal regime.} 
\label{Fig_energy}
\end{figure}

This evaluates, with a continuous function across the three cases of 
$r>1$, $r=1$, $r<1$, to 
\bea 
E(r>1)&=&\frac{g}{24\pi}\,\left[\frac{\pi}{4}-\frac{r}{2(r^2-1)}\right.\\ 
&\,&\qquad\left.-\frac{r(2-r^2)}{2(r^2-1)^{3/2}}\,\left(\sin^{-1}\frac{1}{r}-\frac{\pi}{2}\right)\right]\notag\\ 
E(r=1)&=&\frac{\kp(3\pi-8)}{288\pi}\\ 
E(r<1)&=&\frac{g}{24\pi}\,\left[\frac{\pi}{4}+\frac{r}{2(1-r^2)}\right.-\\ 
&\,&\qquad\left.-\frac{r(2-r^2)}{2(1-r^2)^{3/2}}\,\ln\frac{1+\sqrt{1-r^2}}{r}\,\right]\ .\notag
\eea 
Limiting cases are $E(r\gg1)=g^3/(192\kp^2)$ and $E(r\ll1)=g/96$. 
Figure~\ref{Fig_energy} plots the radiated total energy.

\section{Distributions}
\label{sec:distributions}

\subsection{Spectral Distribution}

The spectral distribution of the electron's radiation is calculated with the help of the formula (see \cite{Jackson:490457} and also Eq.~26 of \cite{Ievlev:2023inj})
\begin{equation}
    \frac{\diff I(\omega)}{\diff \Omega} = 
        \frac{\omega^2}{16 \pi^3} \sin^2\theta \, \abs{   j_z(\omega, k_z ) }^2 \ ,
\label{dIdOmega_definition}
\end{equation}
where $k_z = \omega \cos\theta$, and the Fourier transform of the current is defined as
\begin{equation}
\begin{aligned}
	j_z (\omega, k_z) 
		&= e\, \int\limits_{- \infty}^{\infty} \diff{t} \, \dv{z}{t} \, e^{-i (\omega t - k_z z(t)) } \\
		&=  e\,  \int\limits_{- \infty}^{\infty} \diff{z} \, e^{-i (\omega t(z) - k_z z) } \ .
\end{aligned}
\label{jz_fourier_transform_definition}
\end{equation}

Using the trajectory from Eq.~\eqref{SP_traj} gives the spectral distribution  
\begin{equation}
	\frac{\diff I }{\diff \Omega} 
		= \frac{e^2 \omega^2 \sin^2\theta}{16 \pi^3 \kappa^2}\,   e^{- \frac{\pi \omega (1 + \cos\theta )}{2 \kappa} } \, 
			\abs{ K_{\frac{i \omega (1 + \cos\theta )} {2\kappa}}\left( \frac{\omega}{g} \right)  }^2\ ,
\label{dIdOmega}
\end{equation} 
where $K$ is a modified 
Bessel function of the second kind. 

\begin{figure}[htbp]
\centering 
\includegraphics[width=\columnwidth]
{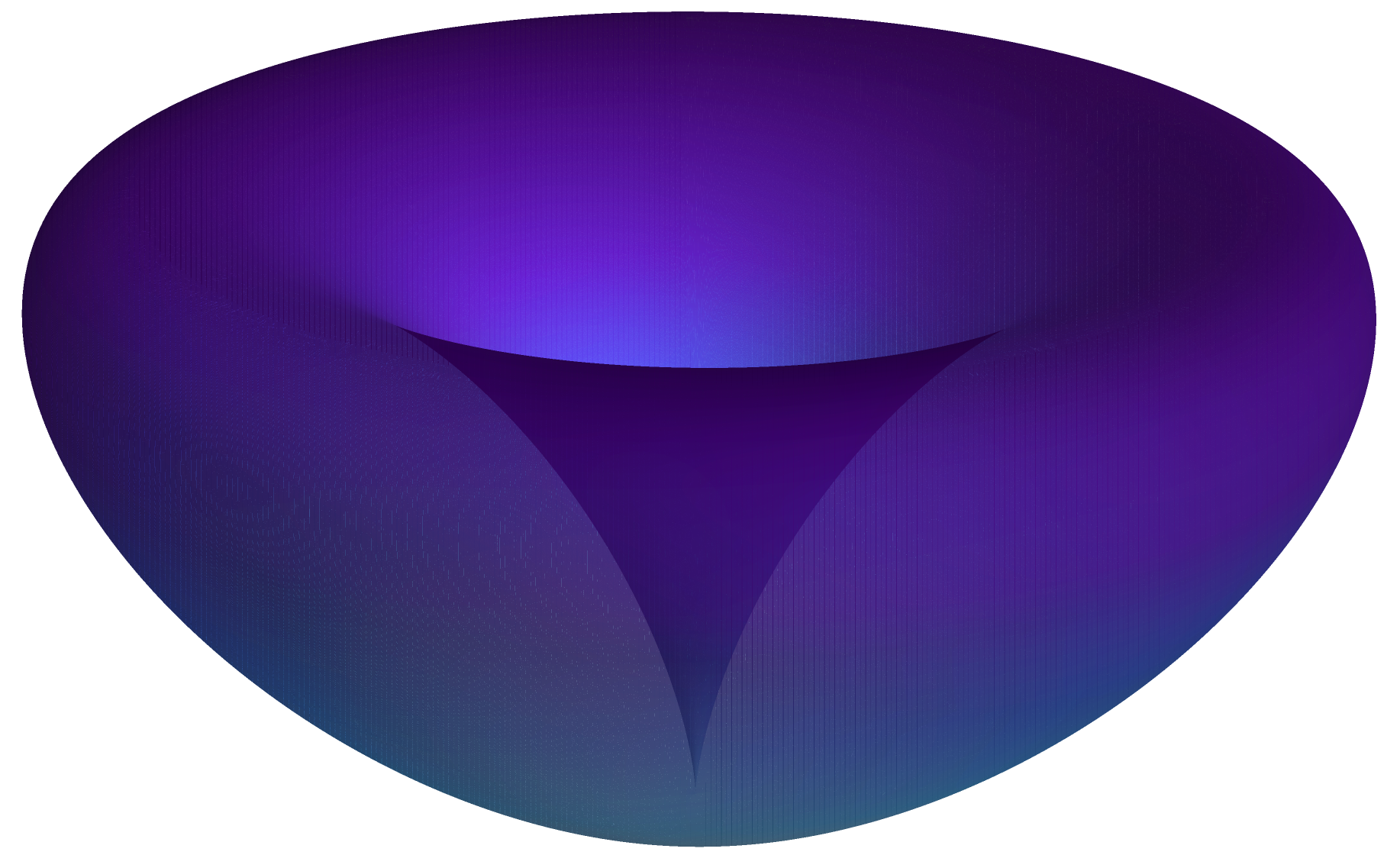}
 \caption{The photon spectral distribution $\diff{I}/\diff{\Omega} = \diff{E}/\diff{\omega}\diff{\Omega}$, Eq.~(\ref{dIdOmega}), emitted by the electron moving downward with a Schwarzschild-Planck trajectory is shown, here for $\kappa = \omega = 1$ 
 and $g = 100$. 
 } 
\label{Fig_specdis}
\end{figure}

\begin{figure}[htbp]
\centering 
\includegraphics[width=\columnwidth]{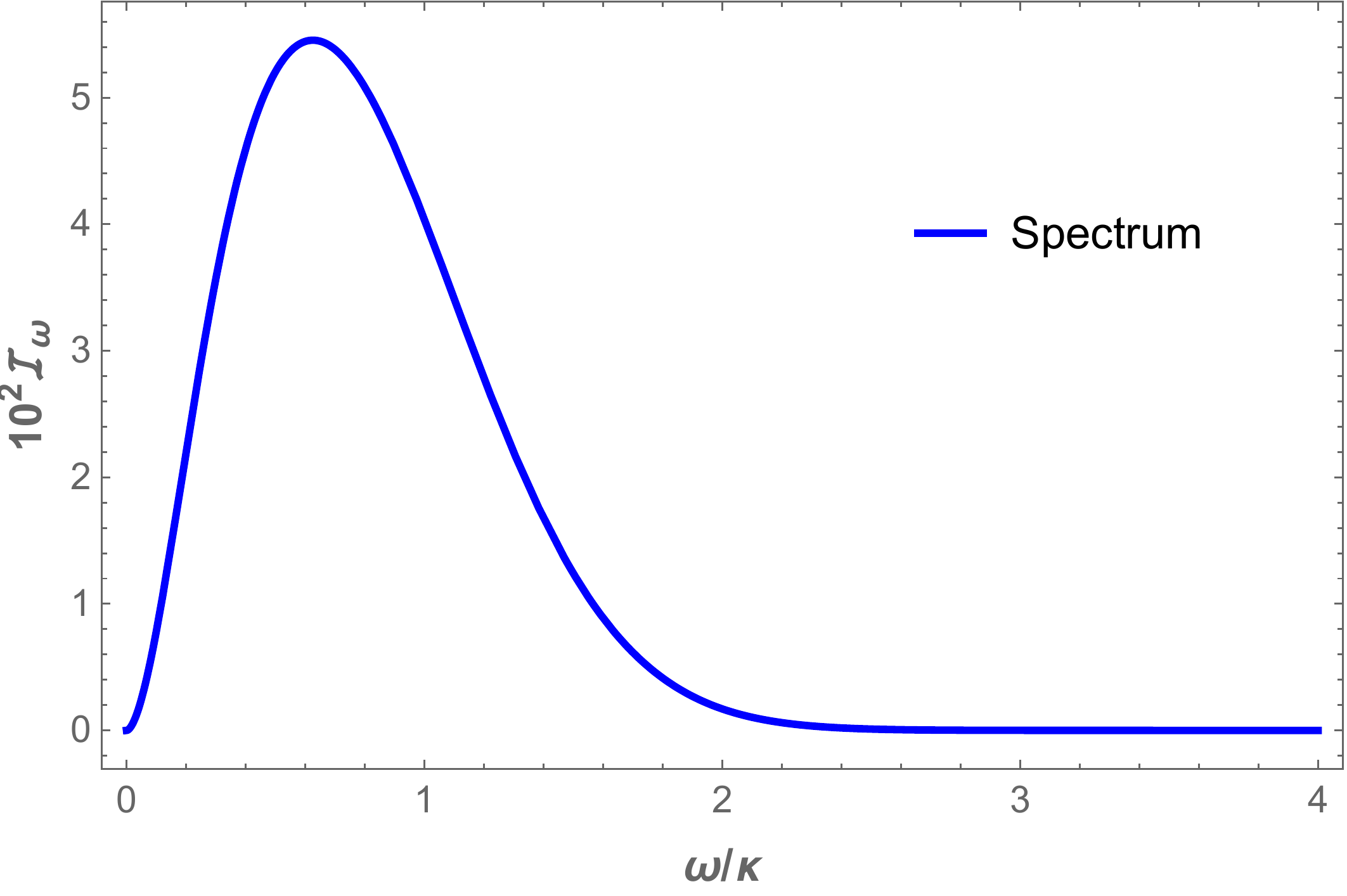}
 \caption{The photon spectrum $I(\omega) = \diff{E}/\diff{\omega}$ emitted by the electron moving along the Schwarzschild-Planck trajectory is shown 
 for the same parameters as Fig.~\ref{Fig_specdis}. 
 } 
\label{Fig_spectrum}
\end{figure}

Figure~\ref{Fig_specdis} plots this 
spectral distribution. Note the lack of 
emission in the $\theta=0$ direction and 
the strong emission in the transverse, 
$\theta=\pi/2$ direction. We also expect 
exponential suppression at high frequency. Integration over solid angle $\diff{\Omega} = \sin\theta \diff{\theta}\diff{\phi}$ gives the spectrum $I(\omega)$, plotted in 
Figure~\ref{Fig_spectrum}.

\subsection{Power Distribution} 

\begin{figure}[htbp]
\centering 
\includegraphics[width=\columnwidth]{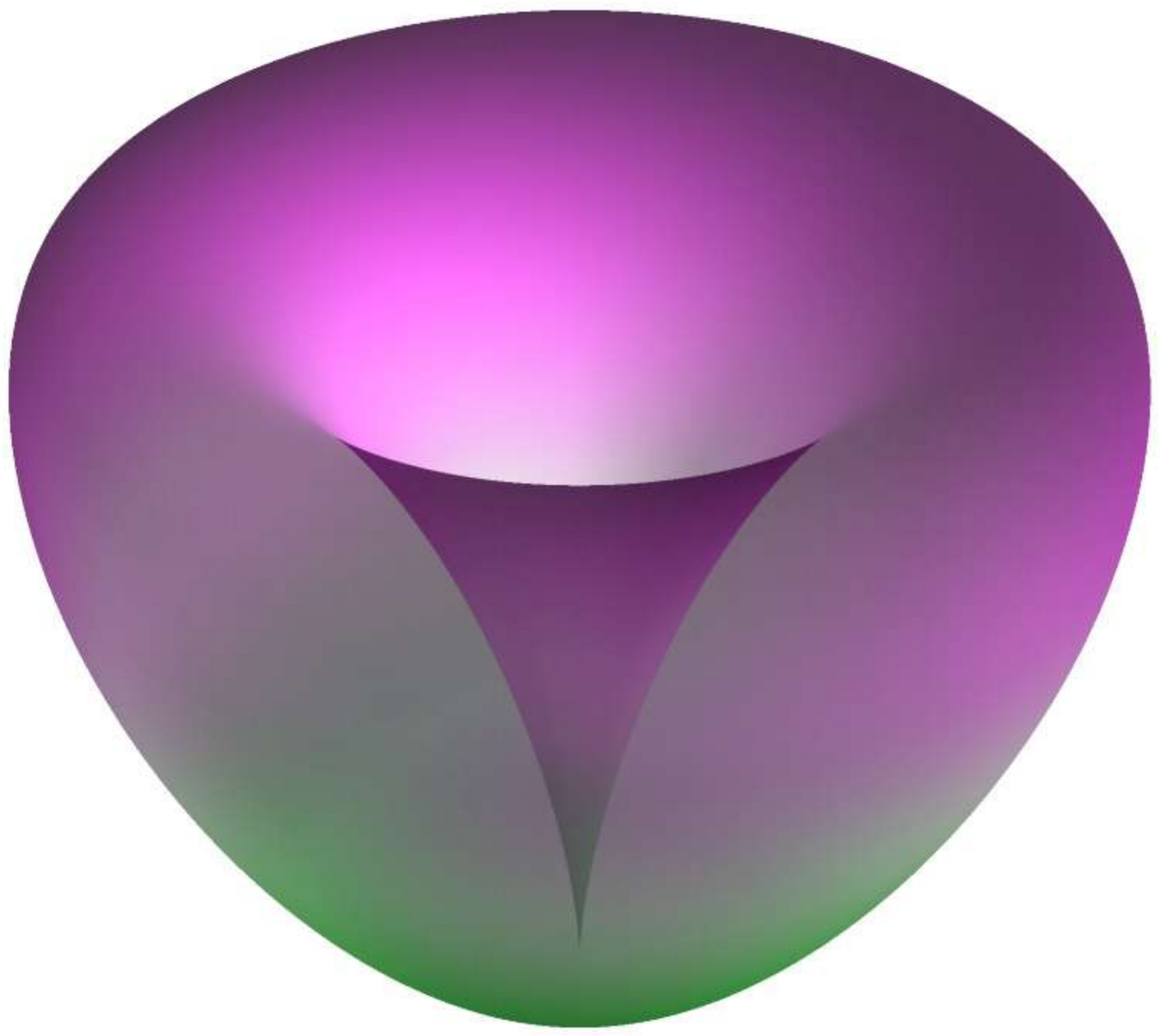}
 \caption{The photon power distribution $\diff{P}/\diff{\Omega} = \diff{E}/\diff{t}\diff{\Omega}$, 
Eq.~(\ref{powdis}), emitted by the electron moving downward with a Schwarzschild-Planck trajectory. Here $\kappa = 1$ and $z=2$ (recall that $z$ defines a specific time $t$) and $g = 100$. 
} 
\label{Fig_powdis}
\end{figure}

A key advantage of the classical approach is the analytic tractability of spacetime and spectral domains.   For instance, the implicit time-dependent power distribution (computed as a function of $z$) is found using Eq.~(\ref{SP_traj}) with straightforward vector algebra \cite{Griffiths:1492149} (see also 
e.g.\ the procedure in \cite{Good:2019aqd}), 
\be 
\frac{\diff{P}}{\diff{\Omega}} = \frac{g^4 \kappa ^4 \sin ^2\theta\, \sinh ^2 2 \kappa  z}{\pi ^2 (g+2 \kappa  \cosh 2 \kappa  z) (g+g \cos\theta +2 \kappa  \cosh 2 \kappa  z)^5}, \label{powdis} 
\ee 
Figure~\ref{Fig_powdis} shows the power distribution $\diff{P}/{\diff{\Omega}}$.  
Its integration over solid angle $\diff{\Omega}$ gives the Larmor power, Eq.~(\ref{eq:power}).  Further integration over time gives the total energy 
\be 
E = \int\frac{\diff{P}}{\diff{\Omega}} \diff{\Omega}\frac{\diff{t}}{\diff{z}}\diff{z}\ . 
\ee

\section{Thermality and asymptotic limits} 
\label{sec:thermality}

\begin{figure*}[t]
	\subfloat[]{\label{fig:distr_theta0}\includegraphics[width=0.99\columnwidth]{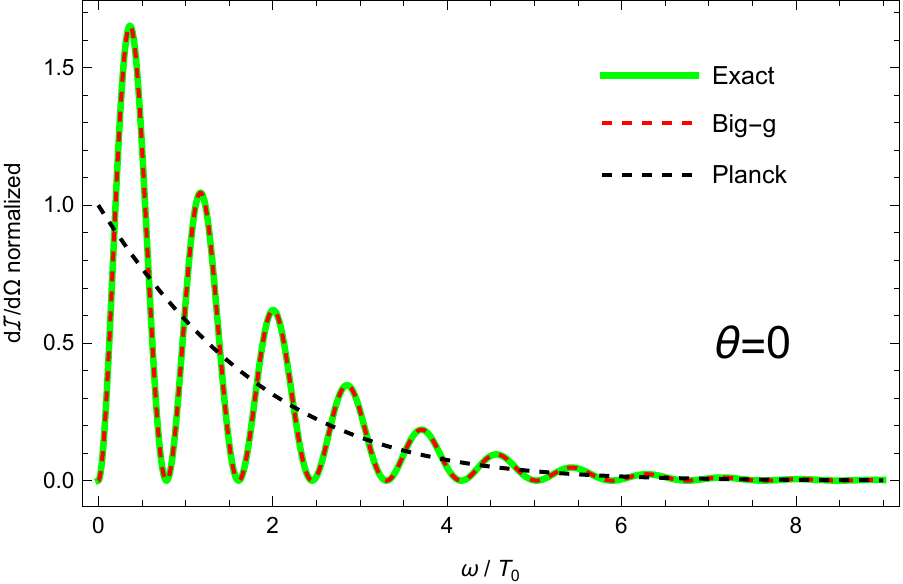}}
	\qquad
	\subfloat[]{\label{fig:distr_theta_PIover2}\includegraphics[width=0.95\columnwidth]{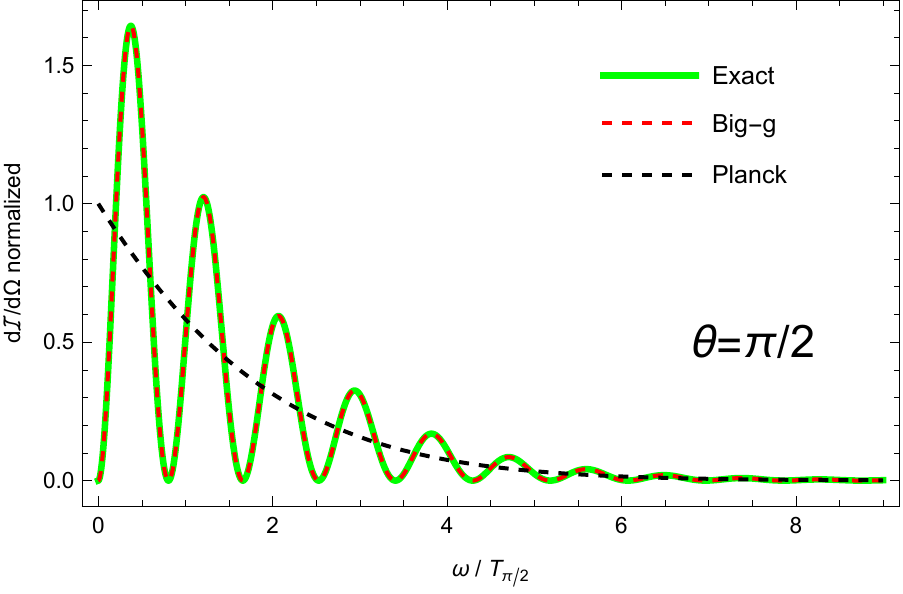}}
	\\
	\subfloat[]{\label{fig:distr_theta_PiminusEps}\includegraphics[width=0.99\columnwidth]{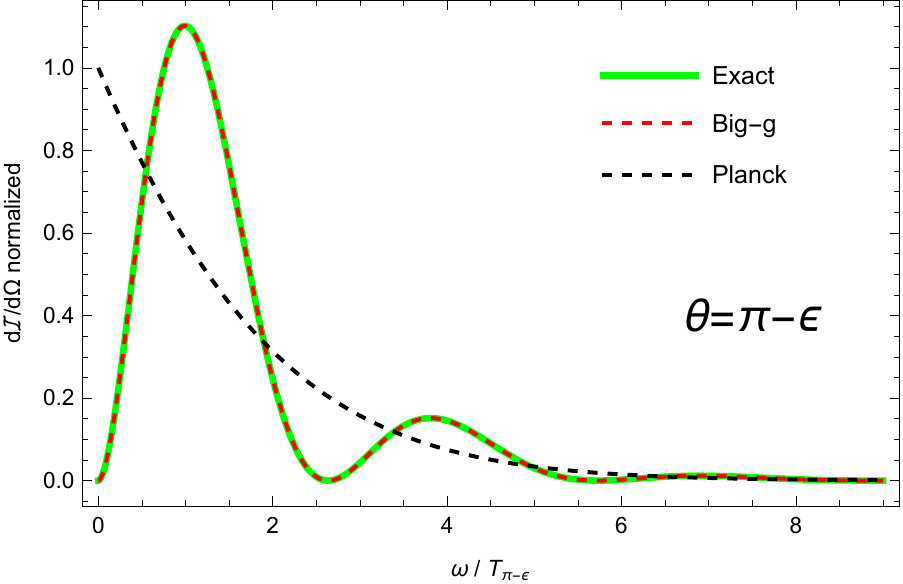}}
	\qquad
	\subfloat[]{\label{fig:distr_theta_Pi}\includegraphics[width=0.99\columnwidth]{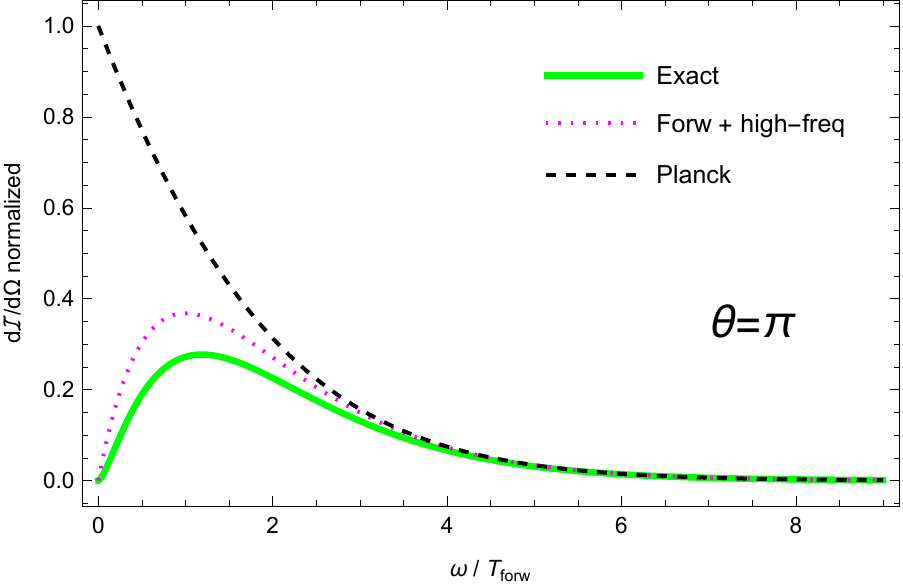}}
\caption{
The spectral distribution, normalized as discussed in the text, is plotted for various directions $\theta$ of 
  emission (as labeled in each panel), here for $g/\kappa = 10^{10}$. 
Top left panel shows the backwards 
direction, top right is transverse to the 
motion, bottom right is exactly forwards, 
and bottom left is nearly but not exactly 
forwards ($\epsilon=50\sqrt{\kappa / g}=5 \cdot 10^{-4}$). 
		The thick green line is the result Eq.~\eqref{dIdOmega}, the black dashed curve is the Planck factor Eq.~\eqref{planck_distr}, 
both normalized by a factor such that the Planck curve 
has the value 1 at the low-frequency asymptote.
		The red dashed line on panels (a)-(c) is the big-$g$ asymptotic Eq.~\eqref{dIdOmega_big_g}, while the magenta dotted 
line on panel (d) represents the high-frequency, forward direction asymptotic Eq.~\eqref{dIdOmega_blue_forw}.
		The spectral distribution away from the forward direction looks very similar at different angles when 
normalized and with the frequency axis scaled by a corresponding temperature $T_\theta$. 
	}
\label{fig:distr}
\end{figure*}

The spectral distribution Eq.~\eqref{dIdOmega} 
has a rich structure and here we consider 
various limiting cases and their relation to 
the Planck distribution denotes thermal 
equilibrium.

\subsection{ Planck Spectrum for Large $g$ }

Recall that the two length parameters entering 
the trajectory, $\kp^{-1}$ and $g^{-1}$, were 
connected in the original Schwarzschild-Planck 
case to the black hole Schwarzschild radius (and 
hence mass) and the Planck length, respectively. 
In that case, we expect $g\gg\kp$. In the present 
case there is an additional subtlety when 
considering the angular distribution in that 
$\kp/(1+\cos\theta)$, not just $\kp$, enters the 
spectral distribution. 

We introduce the quantities 
\begin{equation}
	\kappa_\theta = \kappa \, \frac{ 2 }{ 1 + \cos\theta } \ , \quad
	r_\theta = \frac{ \kappa_\theta }{g}\ . 
\label{kappa_r_theta_definition}
\end{equation} 
In the limit $ r_\theta \ll 1$ 
the spectral distribution Eq.~\eqref{dIdOmega} reduces to a Planck form with oscillations, 
\bea 
	\frac{1}{ e^2 \sin^2\theta } \frac{\diff I}{\diff \Omega} \Bigg|_{r_\theta \ll 1} 
		&=& \frac{ 1 }{ 8 \pi^2 \kappa (1 + \cos\theta )} 
			\, \frac{ \omega }{ e^{ 2\pi \omega / \kappa_\theta } -1 }\notag\\ 
&\,&\quad\times\left[1 - \cos\varphi(\omega, \kappa_\theta , g) \right] \ . 
\label{dIdOmega_big_g} 
\eea 
See Appendix~\ref{sec:asympt_technicalities} for the derivation.
A far-away observer situated at an angle $\theta$ 
(where $\theta=0$ means the observer is behind the electron's direction of motion) 
would see the temperature
\begin{equation}
	T_\theta = \frac{ \kappa_\theta }{2 \pi}\ .
\label{temperature_theta}
\end{equation}
The phase
\begin{equation}
	\varphi(\omega,\kappa_\theta,g) = \frac{2\omega}{\kappa_\theta} \ln\frac{\omega}{2g} - 2 \operatorname{Arg} \left\{ \Gamma(1 + i \omega/\kappa_\theta) \right\}
\label{phase_big_g}
\end{equation}
represents rapid oscillations (recall 
$r_\theta\ll1$) with frequency $\sim (2/\kappa_\theta) \ln(g / \kappa_\theta)$. 
Note that $g$ only enters in the frequency 
of the oscillations, not their amplitude, which 
simply ranges over 0 to 2. 
The oscillations arise from the presence of a 
second length scale (i.e.\ $1/g$); see the 
discussion in \cite{Fernandez-Silvestre:2022gqn} about their relation to 
interference effects between two effective 
horizons, even if the horizons don't manifest 
physically. 
Averaging out the oscillations, we obtain a pure Planck distribution 
\begin{equation}
	\frac{1}{ e^2 \sin^2\theta } \frac{\diff I}{\diff \Omega} \Bigg|_{\substack{r_\theta \ll 1 \\[2pt]  +\text{avg} } }
		\approx \frac{ 1 }{ 8 \pi^2 \kappa (1 + \cos\theta )} 
			\, \frac{ \omega }{ e^{ \omega / T_\theta } -1 } \ . 
\label{planck_distr}
\end{equation} 
The appearance of the Planck factor is consistent with known results for the corresponding mirror \cite{GoodMPLA,Good:2019tnf}, see Sec.~\ref{sec:electron-mirror} here for the electron-mirror dictionary.

Figure~\ref{fig:distr} plots the spectral distribution, 
normalized by $e^2\sin^2\theta/[8\pi^3(1+\cos\theta)^2]$, as well as a Planck distribution 
with temperature $T_\theta$, normalized the same way. 

For example, the limit $\theta \sim 0$ (where 
the observer is behind the electron, seeing it 
redshift as it recedes) gives the temperature $T_0 = \kappa / 2 \pi$, see Fig.~\ref{fig:distr_theta0}.
An \textquote{orthogonal} observer at $\theta = \pi/2$ would see the electron twice as hot, $T_{\pi/2} = \kappa / \pi$, see Fig.~\ref{fig:distr_theta_PIover2}.

At big $g$ this approximation remains valid in a wide region of angles $\theta \in [0, \pi - \epsilon_\star]$, where $\epsilon_\star = O(\sqrt{\kappa / g})$, so that 
$r_\theta\ll1$ stays valid. 
Throughout all of this range oscillations follow the smooth Planck curve, and their average is the 
Planck distribution. 

In a narrow region close to the forward direction 
$(\pi - \theta) \lesssim \kappa / g $, the condition 
$r_\theta\ll 1$ no longer holds. 
The oscillations become increasingly spread out, and the averaging to thermal gets steadily worse, see Fig.~\ref{fig:distr_theta_PiminusEps}. 
We treat this forward region next.

\subsection{Wien's Law in the Blueshift-Forward Limit} 

When $\theta \sim \pi$ (the observer is in front of the electron), 
the Bessel function in Eq.~\eqref{dIdOmega} reduces to $K_0$, giving 
\begin{equation}
	\frac{1}{ e^2 \sin^2\theta } \frac{\diff I}{\diff \Omega} \Bigg|_{\text{forw}} 
		\approx \frac{ \omega^2 }{16 \pi^3 \kappa^2 }  
			\ \abs{ K_{ 0 }\left( \frac{\omega}{g} \right)  }^2 
\label{dIdOmega_blue}
\end{equation}
In this case, a high-frequency asymptotic $\omega \gg g$ can be easily found, 
\begin{equation}
	\frac{1}{ e^2 \sin^2\theta } \frac{\diff I}{\diff \Omega} \Bigg|_{\substack{ \text{high-freq} \\[2pt]  + \text{forw}  } } 
		\approx \frac{ g }{ 32 \pi^2 \kappa^2 } \   \omega e^{- \omega/T_\text{forw} }  \ , 
\label{dIdOmega_blue_forw}
\end{equation} 
where $T_\text{forw}=g/2$. 
This equation is Wien's law with the temperature $T_\text{forw}$ (without oscillations). 
However, there is no Planck factor, so one can call this regime \textquote{thermal} only in a high-frequency sense, see Fig.~\ref{fig:distr_theta_Pi}.

If the angle $\theta$ is not quite at zero, the formula for the temperature $T_\text{forw}$ receives corrections, see Eq.~\eqref{temperature_rtheta_greaterthan_1} and Fig.~\ref{fig:temperature}.

\subsection{ Infrared Limit }

At low frequencies $\omega \to 0$ the spectral distribution Eq.~\eqref{dIdOmega} has an asymptotic form
\begin{equation}
	\frac{\diff I(\omega)}{\diff \Omega} 
		\approx \frac{e^2 \omega^2 \sin^2\theta}{16 \kappa^2 \pi^3}   
					\left( \ln\frac{\omega}{2 g} + \gamma \right)^2 \ ,
\label{dIdOmega_asympt_IR}
\end{equation}
where $\gamma$ is the Euler--Mascheroni constant (see Appendix~\ref{sec:asympt_technicalities} for the details).
Note that the exact spectral distribution vanishes as $\omega \to 0$, and one should use the averaged formula Eq.~\eqref{planck_distr} with care in the IR. A vanishing spectrum in the IR leads to finite particles, which will be computed in the following section.

\subsection{Finite Particles} 

\begin{figure}[htbp]
\centering 
\includegraphics[width=\columnwidth]{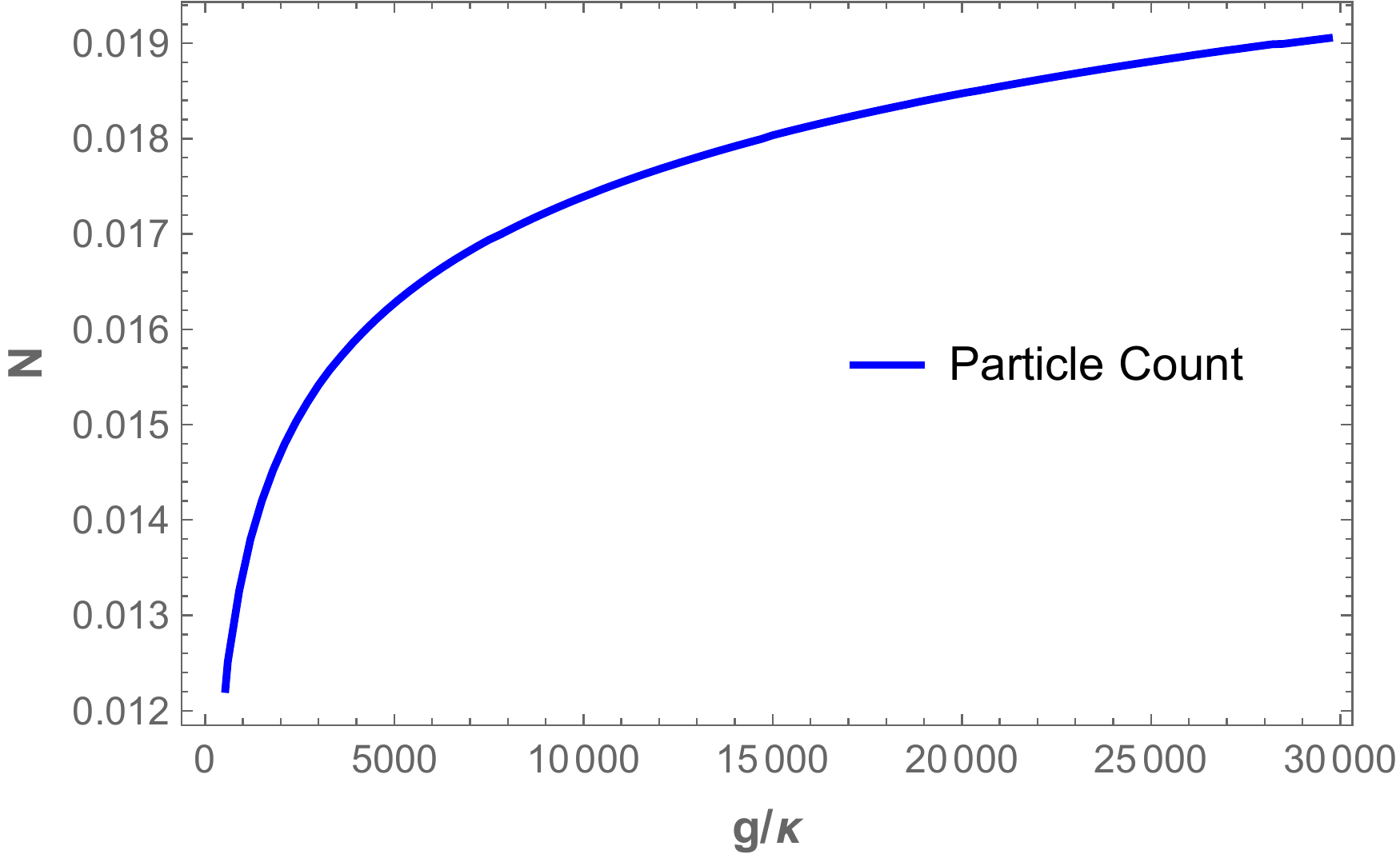}
 \caption{Photon count, Eq.~(\ref{particles}), emitted by the electron moving along the Schwarzschild-Planck trajectory 
 is plotted vs $g/\kp$. Recall that $N_\gamma\propto e^2 = 4\pi \alpha_{\textrm{fs}} \approx 0.092$, in units where $\hbar = c = \epsilon_0 = \mu_0 = 1$. 
 } 
\label{Fig_particles}
\end{figure}

The single electron moving along the Schwarzschild-Planck trajectory radiates a finite number of photons. The particle spectrum $\diff{N}_\gamma(\omega)/\diff{\omega}$ is, using Eq.~(\ref{dIdOmega}),
\be 
\frac{\diff{N}_\gamma}{\diff{\omega}} = \int \frac{1}{\omega}\frac{\diff{I}}{\diff{\Omega}}\diff{\Omega}\ . 
\ee
Integrating over all the frequencies $\omega$ gives the total number of photons emitted $N_\gamma$, 
\be 
N_\gamma = \int_0^\infty \int_0^{2\pi} \int_0^\pi \frac{1}{\omega}\frac{\diff{I}}{\diff{\Omega}}\sin\theta\diff{\theta}\diff{\phi}\diff{\omega}\ . \label{particles} 
\ee
Figure~\ref{Fig_particles} plots the total particle count as 
a function of $g/\kappa$. This is a key difference between the two types of asymptotic inertial motions: the finite particle count of asymptotically zero-velocity trajectories is a signature property in contrast to the infinite soft particle production of asymptotically non-zero sub-light velocity trajectories.  We note that finite energy, entropy, and particle count are useful qualities for correspondence with realistic experiments and unitary physical conditions.

\section{Radiant electron-mirror correspondence}
\label{sec:electron-mirror}

We now return to the connection between classical electrodynamics (a point charge moving in 3+1d) and quantum field theory (dynamical boundary condition for a scalar field in 1+1d).

Let us start with reviewing some background regarding this correspondence (see e.g.\ \cite{Birrell:1982ix}).
In the moving mirror setup, we consider a massless scalar field in 1+1-dimensional spacetime \cite{DeWitt:1975ys,Davies:1976hi,Davies:1977yv}.
If the spacetime were actually Minkowski, the vacuum is trivial, and in the absence of sources, there is no radiation.
However, a moving mirror is a dynamical boundary condition \cite{Good:2021asq}, and this causes perturbation of the vacuum and creation of scalar field quanta, 
as the dynamical Casimir effect. 
An observer sees this as radiation. 

The total energy radiated as a result of this dynamical Casimir effect is (see e.g.\ \cite{walker1985particle})
\begin{equation}
	E_\text{tot}^\beta = \int\limits_0^\infty dp \int\limits_0^\infty dq \ p\; |\beta_{pq}|^2  \ .
\label{Etot_betas}
\end{equation}
Here $|\beta_{pq}|^2 = |\beta^R_{pq}|^2 + |\beta^L_{pq}|^2$, and $\beta^{R,L}_{pq}$ are the beta Bogolyubov coefficients for the right and left sides of the mirror \cite{Parker:2009uva} (see also \cite{Zhakenuly:2021pfm,Good:2022gvk,Good:2023hsv}). 
Frequencies $p$ and $q$ refer to outgoing and incoming plane waves, see e.g.\ \cite{Fabbri}.

\subsection{Short Review of Previous Results} 

Following Nikishov-Ritus \cite{Nikishov:1995qs} (see also \cite{Ritus:1999eu,Ritus:2002rq,Ritus:2003wu,Ritus:2022bph}), the authors of \cite{Ievlev:2023inj,Good:2023ncu} related the beta Bogolyubov coefficients $|\beta_{pq}|^2$ to a certain Lorentz scalar characteristic of the electron moving in 3+1d. 
The formula stated in \cite{Ievlev:2023inj} is 
\begin{equation}
    \frac{|j_\mu (\omega, k_z)|^2 + |j_\mu (- \omega, k_z)|^2}{(2\pi)^2 e^2 } = |\beta_{pq}|^2 \ ,
\label{beta_j_general}
\end{equation} 
(note that $|j_\mu (-\omega, k_z)|^2 = |j_\mu (\omega, -k_z)|^2$) 
where $2p = \omega + k_z$, $2q =\omega - k_z$, and the scalar $|j_\mu (\omega, k_z)|^2$ is defined as 
\begin{equation}
	|j_\mu (\omega, k_z)|^2 \equiv |\bd{j}(\omega, k_z)|^2 - |\rho(\omega, k_z)|^2 \ .
\label{j_scalar}
\end{equation}
Here $\bd{j}$ and $\rho$ are the Fourier-transformed three-current and charge density.
The correspondence propagates to the 
total energy radiated by an electron 
through the equivalence \cite{Ievlev:2023inj,Good:2023ncu} of 
\begin{equation}
\begin{aligned}
	E_\text{tot}^e
		&= \int\limits_0^\infty d\omega \int d\Omega\, \frac{\diff I(\omega)}{\diff \Omega} \ ,\\ 
\end{aligned}
\label{Etot_pointcharge}
\end{equation} 
and 
the total energy of radiation coming from a mirror that moves on the same trajectory,
\begin{equation}
	E_\text{tot}^\beta = \int\limits_0^\infty dp \int\limits_0^\infty dq \ p\; |\beta_{pq}|^2  \ . 
\label{Etot_betas_1}
\end{equation}

\subsection{Refined Recipe Derivation} 

Now we derive a more refined version of this correspondence, 
given a full derivation of 
Eq.~\eqref{beta_j_general} as opposed to the previous 
more heuristic motivation (and check of the final result). 

We begin by writing down the formula for the beta Bogolyubov coefficients.
For the right side of the mirror,
(see Eq.~2.23 of \cite{good2013time}, Eqs.~77-80 of \cite{Good:2016atu}, or \cite{Fabbri,Birrell:1982ix}), 
\begin{equation}
	\beta_{p q}^R=\frac{1}{4 \pi \sqrt{pq}} \int\limits_{-\infty}^{v_0} \diff v\, e^{-i q v-i p f(v)}\left(q- p\, \dv{f(v)}{v} \right) \ .
\label{betaR_def}
\end{equation}
In this equation, the mirror's trajectory is defined as $u = f(v)$ in terms of the lightcone coordinates
\begin{equation}
	u = t - z \,, \quad 
        v = t + z \ . 
\end{equation}
This is equivalent to defining the trajectory in the form $z = z(t)$.
Here $v_0$ is the position of the mirror's horizon; for a mirror without a horizon (like the Schwarzschild-Planck case we focus on in this paper) one should set $v_0 = \infty$.

Making a change of variables in Eq.~\eqref{betaR_def} we 
pass from integration over $v$ to the integration over $t$ according to
\begin{equation}
	u = t - z(t) \ , \quad 
        v = t + z(t) \ .
\end{equation}
Moreover, we substitute the frequencies $p, q$ by the electron-related frequency as
\begin{equation}
    p + q = \omega \,, \quad
    p - q = \omega \cos\theta \ .
\label{pq_definition_2}
\end{equation}
The ingredients of Eq.~\eqref{betaR_def} become under this transformation
\begin{equation}
\begin{aligned}
	\diff v &= (\dot{z}(t) + 1) \diff t \ , \\ 
	q- p \frac{d f(v)}{d v} &= \omega\, \frac{\dot{z}(t) - \cos\theta }{ \dot{z}(t) + 1 } \ , \\
	q v + p f(v) &= \omega \, (t - z(t) \, \cos\theta) \ .
\end{aligned}
\end{equation} 
Now we can easily write down a transformed version of Eq.~\eqref{betaR_def}:
\begin{equation*}
\begin{aligned}
	\beta_{p q}^R 
		&= \frac{1}{2 \pi \sin\theta } \int\limits_{-\infty}^{\infty} \diff t\, (\dot{z}(t) - \cos\theta)\, e^{- i \omega (t - z \cos\theta) } \\
		&= \frac{1}{2 \pi e \sin\theta }\, [ j_z (\omega, \omega\cos\theta) - \rho(\omega, \omega\cos\theta) \cos\theta ]\ . \\ 
\end{aligned}
\end{equation*} 
where the charge and current densities are 
\be 
\rho=e\,\delta(z-z(t))\,,\quad j_z=e\dot z\,\delta(z-z(t))\ . 
\ee 
Using the charge conservation law
\begin{equation} 
	\omega \rho (\omega, \bd{k}) - \bd{k} \cdot \bd{j} (\omega, \bd{k}) = 0 \ , 
\label{cont_eq}
\end{equation}
we arrive at the final answer
\begin{equation}
	\beta_{p q}^R = \frac{\sin\theta }{2 \pi e  }\ j_z (\omega, \omega\cos\theta)\ . 
\label{recipe-betaR}
\end{equation}
This gives a refined form of the correspondence.

The beta Bogolyubov coefficients for the left side of the mirror are related to the right side by a parity flip, $\beta^L_{pq} = \beta^R_{qp}$. 
One can repeat the derivation or realize the parity 
flip is equivalent to $\theta\to\pi-\theta$. Either way 
one easily obtains the previously stated correspondence Eq.~\eqref{beta_j_general} written for both sides of the mirror with $|\beta_{pq}|^2 \equiv |\beta^R_{pq}|^2 + |\beta^L_{pq}|^2$.

\subsection{Practical Significance}

Although the newly derived correspondence is interesting and intriguing in and of itself, it also has a potential for applications.
Taking the absolute value squared of Eq.~\eqref{recipe-betaR} and comparing with Eq.~\eqref{dIdOmega_definition} we can relate the mirror to the electron's spectral distribution: 
\begin{equation}
\begin{aligned} 
\frac{1}{e^2 \omega^2} \, \frac{\diff{I}}{\diff{\Omega}}(\omega,\cos\theta) &= \frac{1}{4\pi} |\beta^R_{pq}|^2 \,,
\\
	p + q = \omega \,, \quad &p - q = \omega \cos\theta \,.
\end{aligned}
\label{recipe_dIdOmega_from_mirror_1}
\end{equation} 
The l.h.s.\ contains only electron quantities, while the r.h.s.\ is in terms of mirror quantities. 
This formula allows one to convert between the radiation from an accelerating boundary (mirror) in 1+1d and an accelerating charge in 3+1d that move on the same rectilinear trajectory.  

We explicitly checked for several trajectories that this correspondence indeed works. 
For example, take the beta Bogolyubov coefficients for the Schwarzschild-Planck mirror
(see e.g.\ Eq.~7 of \cite{Good:2019tnf} and Eq.~3 of \cite{GoodMPLA}):
\begin{equation}
	|\beta_{p q}^R |^2
		= \frac{ pq }{ \pi^2 \kappa^2 (p + q)^2 } \ e^{- \frac{\pi p }{ \kappa} } \,  \abs{ K_{i p/\kappa}\left( \frac{p+q}{g} \right) }^2\ .
\end{equation}
One can immediately see that the recipe Eq.~\eqref{recipe_dIdOmega_from_mirror_1} gives the spectral distribution Eq.~\eqref{dIdOmega}.

For the total energy the correspondence of 
Eq.~\eqref{recipe_dIdOmega_from_mirror_1} 
agrees with the derivation of 
\cite{Good:2023ncu,Ievlev:2023inj} 
that the total energy radiated by an electron is equal to the total energy from both sides  
of the mirror, as can be seen by the 
following. Using 
the electron's total radiated energy formula Eq.~\eqref{Etot_pointcharge} and the correspondence Eq.~\eqref{recipe_dIdOmega_from_mirror_1} we obtain an expression for the total energy from the mirror, 
\begin{equation}
	E_\text{tot}^\beta = \int\limits_0^\infty dp \int\limits_0^\infty dq \ (p + q)\; |\beta^R_{pq}|^2  \,.
\label{Etot_betas_flip}
\end{equation} 
Now recall that a parity flip relates the two sides of the mirror, so  
\begin{equation}
\begin{aligned} 
    E_\text{tot}^\beta 
        &= \int\limits_0^\infty dp \int\limits_0^\infty dq \ p \; |\beta^R_{pq}|^2  
           + \int\limits_0^\infty dp \int\limits_0^\infty dq \ q \; |\beta^R_{pq}|^2 \\
        &= \int\limits_0^\infty dp \int\limits_0^\infty dq \ p \; |\beta^R_{pq}|^2  
            + \int\limits_0^\infty dp \int\limits_0^\infty dq \ q \; |\beta^L_{qp}|^2 \,. \\
\end{aligned}
\label{Etot_check_1}
\end{equation} 
By relabeling the variables $q \leftrightarrow p$ in the last integral we immediately recover Eq.~\eqref{Etot_betas}. 

For convenience, we also provide Table~\ref{tab:limits} giving the 
correspondence between some mirror limits and electron limits. Recall that 
$p=\omega(1+\cos\theta)/2$, $q=\omega(1-\cos\theta)/2$.

\begin{table}[ht] 
\begin{tabular}{l | l }
Mirror & \ Electron   \\ \hline
\rule{0pt}{1.05\normalbaselineskip}moves to the left\ \ & \ moves down $z \to - \infty$  \\
	high-freq.\ $q \gg p$ & \ blueshift-forward $\theta \to \pi$, $q \approx \omega$ \\
	low-freq. $q \ll p$ & \ redshift-recede $\theta \to 0$, $p \approx \omega$
\end{tabular} \\ 
\caption{Correspondence between mirror and electron properties in various limits.} 
\label{tab:limits} 
\end{table}

\section{Other Remarks} 
\label{sec:remarks}
\subsection{\textit{Resistors, accelerating electrons, and black holes}} 

The one-channel Planck curve, Eq.~(\ref{planck_distr}), also appears as the signature form of Nyquist white noise \cite{nyquist} for a resistor. The noise power emitted by a resistor is independent of the resistance (see e.g.\ Eq.~15.17.14 of Reif \cite{reif}), 
\be 
P(\omega) = \frac{\diff{\omega}}{2\pi} \frac{\hbar\omega}{e^{\hbar \omega /k_B T} - 1}\ , 
\ee 
and equivalent to what an antenna would pick up from a three-dimensional black body at the same temperature \cite{Dicke}. Lowering the temperature means lower noise.  This curve is essential to understanding and mitigating the impact of thermal noise on electronic systems at high frequencies or low temperatures, $\hbar \omega \gg k_B T$, due to quantum effects \cite{urick2021fundamentals}. Thus, the one-channel Planck curve's appearance in the classical radiation from an accelerating electron is remarkable. We note that as far as information flow is concerned, a black hole also behaves as a one-dimensional channel \cite{Bekenstein:2001tj,Bekenstein:2003dt,Bekenstein:2001zoa}.

\subsection{\textit{Remarks on electron evaporation}}

 Complete evaporation should address the problem of backreaction. A fixed background approximation ignores the effects of the radiation on the spacetime geometry.  Backreaction effects are expected to play an important role even before reaching the Planck scale when the unknown theory of quantum gravity is expected to dominate \cite{Fabbri}.  The classical thermal Schwarzschild-Planck model of an accelerated electron is an analog for black hole radiance.  The model goes beyond solving for the finite energy emission that is one of the primary consequences of backreaction but also results in finite particle emission, leaving no confusion about the end state of the black hole radiation process: complete evaporation with no left-over remnant \cite{wilczek1993quantum}. 

In the context of quantum theory, an electron can never truly come to a complete stop without considering the uncertainty principle.  At late times, the electron's velocity, Eq.~(\ref{velocity}), becomes arbitrarily small,
\be 
\lim_{t\to\infty}V(t) = 0\,, 
\ee 
as $1/(2\kp t)$, 
and consequently, the wave function becomes more and more spread out over time, in a process not so different to `complete black hole evaporation'; i.e.\ crudely speaking, the black hole `spreads' itself out over time.   
In other words, the de Broglie wavelength of such an electron becomes huge:
\begin{equation}
    \lambda_\text{dB} = \frac{h}{p} \ ,
\end{equation}
where $h$ is Planck's constant and $p$ is the electron's momentum.
Supposing the wave function of the electron becomes infinitely spread out, its magnitude approaches zero, and the probability of finding the electron in a specific region of space becomes very small.

However, the radiation emitted by the point charge in this paper is completely classical.  What are the signatures for classical complete evaporation? The mass of the black hole lessens, converting to radiation energy.  The electron mass never changes.  Nevertheless, the electron `goes away' by traveling to asymptotic infinity, Eq.~(\ref{SP_traj}), 
\be 
\lim_{t\to \infty} z(t) = -\infty\ . 
\ee 
This infinity is an important artifact of the model but is not unique to asymptotic zero-velocity motions.  Consider numerous models, e.g.\ \cite{Good_2015BirthCry}, which are also asymptotic inertial but obtain asymptotic constant velocity, 
\be 
\lim_{t\to\infty}V(t) = V_0 = \textrm{constant}\ , 
\ee 
with $1 >|V_0| > 0$. 
In this case, the de Broglie wavelength stays finite.
These asymptotic constant velocity electrons may also travel to spatial infinity, yet they do not `go away': a residual Doppler shift \cite{wilczek1993quantum} and soft photons \cite{Ievlev:2023inj} signal their eternal presence. How, then, can these classical considerations yield insight into complete evaporation? Consider how the measurement of velocity requires successive measures of position. If an electron is absent, there will be no measurable change in position, which a zero velocity measurement would reflect. 
Consider also that the IR limit $\omega \to 0$ of the classical spectrum $I(\omega)$ is zero for an asymptotically resting electron but non-zero for asymptotic drift.

\section{Conclusions} 
\label{sec:conclusions}
We have found conditions under which an accelerating electron emits thermal radiation. By using the Schwarzschild-Planck trajectory, thermality and asymptotic rest 
(hence finite particle and energy emission) 
are reconciled. Some key takeaways:
\begin{itemize}
    \item A peel acceleration plateau demonstrates a steady-state dynamics connection to the thermal spectrum. The electron accelerates with uniform peel, leading to the emission of thermal radiation.
    \item There is a refined and precise `holographic' duality between the radiation emitted by an accelerating electron in 3+1 dimensions and a moving mirror in 1+1 dimensions.  This recipe explicitly connects the mirror's Bogolyubov coefficients to the electron's spectral distribution.
    \item The temperature arising from calculating the spectral distribution is explicitly related to the acceleration. 
    The necessary two length scales of Schwarzschild-Planck give high-frequency oscillations about a Planck spectrum. 
\end{itemize}
While this paper develops a fully classical radiation model, it establishes critical characteristics expected of complete black hole evaporation (with unitarity, i.e.\ no 
information loss). Several possible extensions related to black holes are of 
interest: 
\begin{itemize}
    \item A modified null-shell collapse could more closely link the properties of black hole radiation and electron radiation. 
    \item A better understanding of the curved geometry associated with the Schwarzschild-Planck spacetime from the spectral insights gleaned here could help develop a geometric perspective of backreaction.
    \item Relaxing the strong constraint of asymptotic rest and investigating asymptotic drift could be an interesting direction, to see whether black hole remnants and asymptotic drifting electrons share radiated particle spectral signatures. 
\end{itemize}
Closer to laboratory experiments, one might apply the fluctuation-dissipation theorem (FDT) to the thermal accelerating electron. The FDT has utilized the 1+1 Planck curve (Eq.\ 4.9 of \cite{Callen:1951vq}) to find the fluctuating voltage of a resistor and other canonical examples like the fluctuating force responsible for Brownian motion.

\begin{acknowledgments} 

Funding comes in part from the FY2021-SGP-1-STMM Faculty Development Competitive Research Grant No.\ 021220FD3951 at Nazarbayev University. This work is supported in part by the Energetic Cosmos Laboratory and in part by the U.S.\ Department of Energy, Office of Science, Office of High Energy Physics, under contract no.\ DE-AC02-05CH11231.  
\end{acknowledgments}

\appendix

\section{Technical Details of the Asymptotic Expansions}
\label{sec:asympt_technicalities} 

In this Appendix we derive the asymptotic formulas presented in the main text, and 
also present some additional technical results.

The first step in this analysis is the asymptotic expansion of the modified Bessel function. We employ the following integral representation, 
\begin{equation}
\begin{aligned}
	K_{i \omega / \kappa_\theta} \left( \frac{\omega}{g} \right) 
		&= \frac{1}{2} \int\limits_{-\infty}^\infty  e^{- \omega f(z) }\, dz \ , \\
	f(z) &= \frac{1}{g} \cosh z + \frac{i}{\kappa_\theta} z \ .
\end{aligned}
\label{Kint}
\end{equation}

\subsection{High-frequency limit}

At high frequencies $\omega$ (i.e.\ roughly speaking 
$\omega\gg g$ or $\omega\gg\kp_\theta$ -- see below) 
one can apply the saddle-point method to the integral in Eq.~\eqref{Kint}.
There are two cases depending on the value of the parameter $r_\theta=\kp_\theta/g$.  

When $r_\theta > 1$, there is only one relevant saddle point: $z_0 = -i \sin[-1](1/r_\theta)$. 
The saddle point approximation works well if
\begin{equation}
    \omega f''(z_0) = \frac{\omega}{g} \sqrt{ 1 - \frac{1}{ r_\theta^2 } }\ \gg 1 \,,
\label{sp_validity_1}
\end{equation}
and under this assumption one obtains 
\begin{equation}
\begin{aligned}
	K_{i \omega / \kappa_\theta} \left(\omega/g \right) 
		&\approx  \sqrt{\frac{ \pi g}{2 \omega \sqrt{1-1/r^{2}_\theta}}} \\
            &\times e^{- \frac{\omega}{g} \left[ \sqrt{1-1/r_\theta^{2}} + (1/r_\theta)\sin^{-1} (1/r_\theta)\right] }  \ .
\end{aligned}
\label{asympt_K_large_rtheta_greaterthan_1}
\end{equation}
Plugging this into the spectral distribution Eq.~\eqref{dIdOmega} gives Wien's law
\begin{equation}
\begin{aligned}
	\frac{\diff I }{\diff \Omega} 
		&\approx \frac{e^2 g  \sin^2\theta}{32 \pi^2 \kappa^2 \sqrt{1 - 1/r_\theta^2} }
			\cdot \omega \, e^{ - \omega / T_{g}}\ , 
\end{aligned}
\label{dIdOmega_asympt_rtheta_greaterthan_1}
\end{equation}
with temperature 
\begin{equation}
	T_{g} = \frac{g}{2} \left( \sqrt{1 - \frac{1}{r_\theta^2} } + \frac{ \pi -    \cos[-1](1/r_\theta) }{ r_\theta }   \right)^{-1} \,.
\label{temperature_rtheta_greaterthan_1}
\end{equation}
Blueshift-forward limit implies $r_\theta \to \infty$, $T_g \to g/2$, and we recover Eq.~\eqref{dIdOmega_blue_forw} in this limit. 

At $r_\theta = 1$ this saddle point merges with another saddle point, and a Stokes-like phenomenon occurs.
In the interval $0 < r_\theta < 1$ two saddle points contribute to the integral in Eq.~\eqref{Kint}: 
$z_{\pm} = - i \pi/2 \pm \ln( 1/r_\theta - \sqrt{ 1/r_\theta^2 - 1 } )$.
The saddle point approximation works well if
\begin{equation}
    \omega \, \abs{ f''(z_\pm) } = \frac{\omega}{\kappa_\theta} \sqrt{ 1 - r_\theta^2  }\ \gg 1 \,.
\label{sp_validity_2}
\end{equation}
The appearance of two saddle points results in an oscillating asymptotic behavior:
\begin{equation}
	K_{i \omega / \kappa_\theta} \left( \omega/g \right) \approx
		\sqrt{ \frac{ 2 \pi \kappa_\theta }{ \omega \sqrt{ 1 - r_\theta^2 } } } 
		\ \cos( \omega \Upsilon / 2 + \pi / 4 )
		\ e^{- \frac{\pi\omega }{2 \kappa_\theta } } \,,
\label{asympt_K_large_rtheta_lessthan_1}
\end{equation}
where the oscillation factor is
\begin{equation}
	\Upsilon = \frac{2}{\kappa_\theta} \left[ \sqrt{ 1 - r_\theta^2 }  + g \ln( 1/r_\theta - \sqrt{ 1/r_\theta^2 - 1 } ) \right] \,.
\end{equation}
Plugging this into the spectral distribution Eq.~\eqref{dIdOmega} gives Wien's law
\begin{equation}
\begin{aligned}
	\frac{\diff I }{\diff \Omega} 
		&\approx \frac{e^2  \sin^2\theta  }{ 8 \pi^2 \kappa (1 + \cos\theta ) \sqrt{ 1 - r_\theta^2 } } \\
            &\times
		\left[ 1 - \cos( \omega \Upsilon  - \pi / 2 ) \right]
		\cdot \omega \, e^{- \omega / T_\theta } \ , 
\end{aligned}
\label{dIdOmega_asympt_rtheta_lessthan_1}
\end{equation}
with temperature $T_\theta$ that exactly coincides with Eq.~\eqref{temperature_theta},
\begin{equation}
	T_\theta = \frac{ \kappa_\theta }{2 \pi} \ .
\label{temperature_theta_appendix}
\end{equation} 

In the limit of large $g$ we generally have $r_\theta \to 0$, and Eq.~\eqref{dIdOmega_asympt_rtheta_lessthan_1} reproduces a high-frequency approximation to the oscillating Planck distribution in Eq.~\eqref{dIdOmega_big_g}. 
Using Stirling's approximation, we checked that the phase Eq.~\eqref{phase_big_g} and the phase of $\cos$ in Eq.~\eqref{dIdOmega_asympt_rtheta_lessthan_1} also agree in the limit of large $g$. 
In the blueshift-forward limit $\theta=\pi$, then 
``large'' $g$ may be small compared to $\kp_\theta=\kp/(1+\cos\theta)$ and so we no longer 
have the limit $r_\theta\to0$. Thus the temperature 
in Eq.~\eqref{temperature_theta_appendix} does not 
actually blow up. 

\begin{figure}[htbp]
\centering 
\includegraphics[width=\columnwidth]{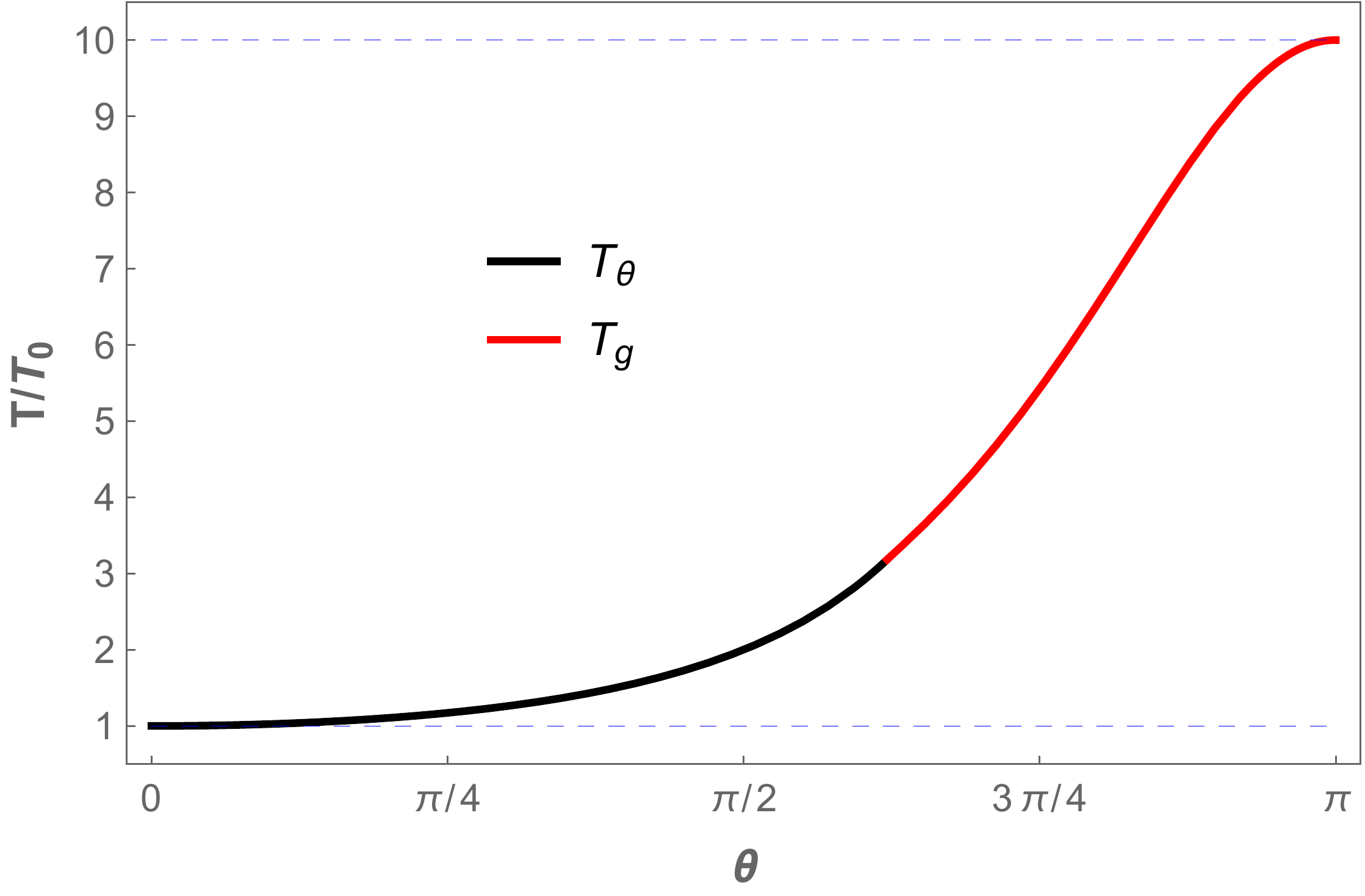} 
\caption{
    Temperature seen by a remote observer positioned at an angle $\theta$ (relevant formulas are Eq.~\ref{temperature_rtheta_greaterthan_1} and Eq.~\ref{temperature_theta_appendix}).
    The temperature is plotted in units of $T_0 = \kappa/2\pi$, and here $\kappa = 2 \pi$, $g = 20$.
    The transition between the redshift-receding $\theta = 0$ limit and the blueshift-forward $\theta = \pi$ limit is indeed smooth, and as expected  
    the temperature rises in this transition. 
}
\label{fig:temperature}
\end{figure}

In fact, there is a smooth transition between 
the branches, with Eq.~\eqref{temperature_rtheta_greaterthan_1} and Eq.~\eqref{temperature_theta_appendix} giving 
the same result at $r_\theta=1$. The full 
behavior is shown in Fig.~\ref{fig:temperature}.

\subsection{Large-$g$ expansion}
\label{sec:appendix_big-g}

We start by analyzing the spectral distribution in the limit $\omega / g \ll 1$. 
The expansion of the modified Bessel function of the second kind is easy in this limit. 
Starting from its definition in terms of the modified Bessel function of the first kind, 
\begin{equation}
	K_{i \alpha} (x)
		= \frac{\pi}{2 \sin (i \alpha \pi) } \left[ I_{- i \alpha} (x) - I_{i \alpha} (x) \right] \,,
\end{equation}
and using the expansion of the $I$-function at small arguments 
 (see e.g.\ Gradshteyn \& Ryzhik 
 Eq.\ 8.445)
\begin{equation}
	I_{ i \alpha} (x) \approx \frac{1}{\Gamma(1 +  i \alpha)} \left( \frac{x}{2} \right)^{ i \alpha} \,, \quad
	x \to 0
\label{I_expansion}
\end{equation}
we can easily arrive at
\begin{equation}
\begin{aligned}
	K_{i \alpha} (x) 
		\approx \frac{\pi}{2 i \sinh (\pi \alpha ) }
			\Big[ 
				&\frac{1}{\Gamma(1 - i \alpha)} \left( \frac{x}{2} \right)^{- i \alpha} \\
				&- \frac{1}{\Gamma(1 + i \alpha)} \left( \frac{x}{2} \right)^{i \alpha} 
			\Big] \ .
\end{aligned}
\end{equation}
Next, using $|a-b|^2 = |a|^2 + |b|^2 - 2 \Re(a^* \cdot b)$ and 
the identity 
\begin{equation}
	\abs{\Gamma(1 + i \alpha)}^2 = \frac{\pi \alpha}{\sinh \pi \alpha } \ , 
\end{equation} 
we find 
\begin{equation}
\begin{aligned}
	&\abs{ K_{i \alpha} (x) }^2 
		= \frac{\pi}{2 \alpha \sinh( \pi \alpha ) } \\
		&\quad \times \Big[ 1 - \cos( 2 \alpha \ln\frac{x}{2} - 2 \operatorname{Arg}(\Gamma(1 + i \alpha)) ) \Big] \,. 
\end{aligned}
\label{Kalpha_abs_approx}
\end{equation}

Plugging this into the spectral distribution Eq.~\eqref{dIdOmega} with $x= \omega / g$, $\alpha = \omega / \kappa_\theta$ yields the formula Eq.~\eqref{dIdOmega_big_g}.

Although we started this derivation under the assumption $\omega / g \ll 1$, the result actually remains valid at large $\omega$ if we have 
$r_\theta = \kappa_\theta / g \ll 1$.
This conclusion is confirmed by the fact that the high-frequency asymptotic result Eq.~\eqref{dIdOmega_asympt_rtheta_lessthan_1} is consistent with the high-frequency limit of Eq.~\eqref{dIdOmega_big_g}.

\subsection{Low-frequency limit}
\label{sec:asympt_lowfrew}

If in addition to $\omega / g \ll 1$ we also assume $\omega / \kappa_\theta \ll 1$, we can get a simple IR asymptotic formula valid for all $\theta$.

In Eq.~\eqref{Kalpha_abs_approx} take $\alpha \ll 1$.
In this limit (see e.g.\ Gradshteyn \& Ryzhik Eq.\ 8.342) 
\begin{equation}
	\ln \Gamma(1 + i \alpha) \approx - i \gamma \alpha \,, 
\end{equation}
and we find $\operatorname{Arg} \Gamma(1 + i \alpha) \approx - \gamma \alpha$. Here $\gamma$ is the Euler--Mascheroni constant.
Then we can expand the $\cos$ and the $\sinh$ in Eq.~\eqref{Kalpha_abs_approx} to obtain
\begin{equation}
	\abs{ K_{i \alpha} (x) }^2
		\approx \left(   \ln\frac{x}{2} + \gamma  \right)^2\ .
\end{equation}
Plugging this with $\alpha=\omega/\kp_\theta$ and $x = \omega / g$ into  the spectral distribution Eq.~\eqref{dIdOmega} we obtain the IR result Eq.~\eqref{dIdOmega_asympt_IR}.
We checked numerically that this approximation indeed works.

\section{Meaning of Large-$g$}

In Sec.~\ref{sec:thermality} we demonstrated that in the limit of large $g$ the electron's spectral distribution acquires the Planck form Eq.~\eqref{planck_distr}, and we may speak of a thermal radiation at all frequencies. 
Here we briefly motivate the reason why large 
$g$ is significant. 

Writing Eq.~\eqref{SP_traj} in dimensionless form as 
\be 
\kp t=-\kp z-\frac{\kp}{g}\sinh 2\kp z \,,
\ee 
immediately shows that small $\kp/g$ gives a longer 
period where the motion is near the speed of light, 
a requirement for getting mirror thermal radiation. 
Moreover, the approach to this 
behavior is exponential, $\sim e^{2\kappa z}$, 
which together yield thermal radiation 
(see \cite{Good_2018,Davies:1977yv} and Carlitz-Willey \cite{CW2lifetime} for further discussion.) 

Recall that in the original Schwarzschild-Planck 
mirror analysis \cite{Good:2020fsw}, 
the conjecture was that the 
new length scale was related to the Planck 
length, with $g\propto 1/l_{\rm Pl}$. Hence we 
indeed expect $g$ to be large compared to other 
scales.

\bibliography{main} 
\end{document}